\documentclass[pre,twocolumn,showpacs]{revtex4}
\usepackage[dvips]{graphicx}
\usepackage{times,epsfig,amssymb}

\begin{document}
\newcommand{\etal}{{\it et al.}}

\title{The $q$-component static model : modeling social networks}
\author{D.-H. Kim, B. Kahng, and D. Kim}
\affiliation{School of Physics and Center for Theoretical Physics,
Seoul National University, Seoul 151-747, Korea}
\date{\today}

\begin{abstract}
We generalize the static model by assigning a $q$-component weight
on each vertex. We first choose a component $(\mu)$ among the $q$
components at random and a pair of vertices is linked with a color
$\mu$ according to their weights of the component $(\mu)$ as in
the static model. A $(1-f)$ fraction of the entire edges is
connected following this way. The remaining fraction $f$ is added
with $(q+1)$-th color as in the static model but using the maximum
weights among the $q$ components each individual has. This model
is motivated by social networks. It exhibits similar topological
features to real social networks in that: (i) the degree
distribution has a highly skewed form, (ii) the diameter is as
small as and (iii) the assortativity coefficient $r$ is as
positive and large as those in real social networks with $r$
reaching a maximum around $f\approx 0.2$.
\end{abstract}

\pacs{89.65.-s, 89.75.Hc, 89.75.Da}

\maketitle
\section{Introduction}
Recently there have been considerable efforts to understand
complex systems in terms of random graph, consisting of vertices
and edges, where vertices (edges) represent individuals
(acquaintances or their interactions)
\cite{Strogatz01,Albert02,Dorogovtsev02,Newman03a}. In such
complex networks, the emergence of a power-law degree
distribution, $P(k)\sim k^{-\gamma}$, is an interesting feature.
Such networks are called scale-free (SF) networks. To illustrate
such SF behavior, many {\it in silico} models have been
introduced, whose examples include the Barab\'asi and Albert
model~\cite{Barabasi99}, the Huberman and Adamic
model~\cite{Huberman99}, etc. In those models, the number of
vertices grows with time.

The static model~\cite{Goh01,Caldarelli02} is another type of {\it
in silico} model designed to generate SF networks, where the
number of vertices is fixed. Each vertex is indexed by an integer
$i$ ($i=1,\cdots,N$) and assigned its own weight
$w_i=i^{-\alpha}$, where $\alpha$ is a tunable parameter. Next,
two different vertices $(i,j)$ are selected with probabilities
equal to normalized weights, $w_i/\sum_k w_k$ and $w_j/\sum_k
w_k$, respectively, and are connected via an edge unless one
exists already. This process is repeated until $mN$ edges are
present in the system, so that the mean degree is $2m$. Then it
follows that the degree distribution is SF with the exponent
$\gamma = 1+1/\alpha$. Thus, tuning the parameter $\alpha$ in
$[0.5,1)$, we can obtain a continuous spectrum of the exponent
$\gamma$ in the range $2 < \gamma \le 3$, for which the degree
distribution has finite mean and diverging variance. Since the
number of vertices does not grow, one may wonder if this model can
be applied to evolving real world network. However, since the
model network can be easily generated and exhibits little hump in
the degree distribution, it can be useful to study many aspects of
SF networks.

In this paper, we generalize the static model by allowing a
$q$-component weight $(w_i^{(1)}, w_i^{(2)},\dots,w_i^{(q)})$ to
each vertex $i$. We suppose that the $\mu$-th component
$w_i^{(\mu)}$ of a vertex $i$ represents its own weight or fitness
to a subgroup $(\mu)$ ($\mu=1,\dots,q$) in a society. For example,
we suppose that two persons $i$ and $j$ are alumni of a high
school, a subgroup $(\mu)$. They would have different weights
$w_i^{(\mu)}$ and $w_j^{(\mu)}$ in the subgroup $(\mu)$,
determined by their school activities. The person $i$ and another
person $k$ are colleagues in a company, another subgroup $(\nu)$.
They have also different weights, $w_i^{(\nu)}$ and $w_k^{(\nu)}$,
by their positions in the company, the subgroup $(\nu)$. Then the
person $i$ has weights $w_i^{(\mu)}$ and $w_i^{(\nu)}$ in
different subgroups, which are not the same in general. We make an
edge between the pair ($i$,$j$) in one color representing the
subgroup $(\mu)$ and the pair ($i$,$k$) in another color for the
subgroup $(\nu)$. Vertices in the system are connected with edges
in $q$ different colors representing different subgroups.
Subgroups are then connected each other by weak ties as explained
later. Since our society comprises many different subgroups and a
person can be acquainted with other people belonging to diverse
subgroups, this generalized static model is useful for modeling
social networks. Meanwhile, it is noteworthy that our model is
reminiscent of the generalization of the Ising model to the
$q$-component cubic model~\cite{Kim75} in equilibrium spin
systems.

\begin{figure}[b]
\includegraphics[scale=0.22]{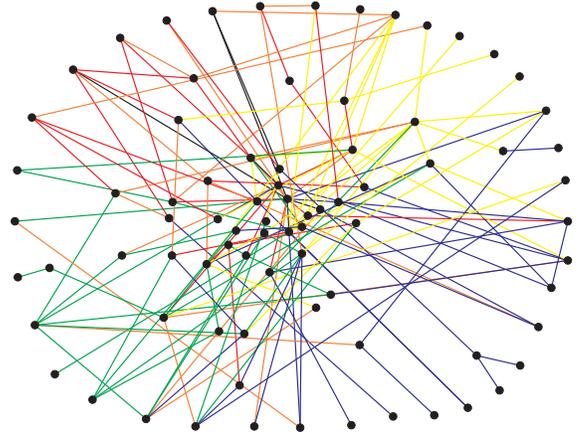}
\caption{A network of the $q$-component static model with
parameters $N=80$, $m=2$, $q=4$ and $f=0.2$. Edges in four colors
(red, yellow, green and blue) are the connections within each
group. Edges in orange are those formed by maximum weights. Edges
in more than two colors are colored in black.} \label{FIG1}
\end{figure}

So far, there have been many attempts to explain the structures
and the properties of social networks \cite{social}. Recently,
Watts \etal~\cite{Watts02} introduced a hierarchical model for
social network. In the model, individuals belong to groups that in
turn belong to groups of groups and so on, creating a tree-like
hierarchical structure of social organization. Here an individual
can belong to more than one group, as a result of which the
distance between two persons is shorter than the ultrametric
distance between them. Such hierarchical model illustrates well
the small-world property of social network as implied in the
Milgram's ``six degrees of separation" \cite{Milgram67}. Another
simple social network model, introduced by Newman and
Park~\cite{Newman03c}, is based on the concept of bipartite graph
~\cite{Newman01} and community structure \cite{Girvan02}. This has
an advantage toward explaining the non-trivial high clustering of
real-world social networks. While our $q$-component model is
similar to the hierarchical model and the community model in the
spirit of dividing people into subgroups, however, we assign
weights to each person for each subgroup, and connections are made
following those weights.

Social network exhibits an interesting feature in the
degree-degree correlation function, different from biological or
information networks. Newman~\cite{Newman02a} studied the
degree-degree correlation in terms of the correlation function
between the remaining degrees of the two vertices on each side of
an edge, where the remaining degree means the degree of that
vertex minus one. He introduced the assortativity coefficient $r$,
defined as
\begin{equation}
r={1\over \sigma_{q}^2}\sum_{j,k}jk (e_{jk}-q_{j} q_{k}),
\label{r_d}
\end{equation}
where $e_{jk}$ is the joint probability that the two vertices on
each side of a randomly chosen link have $j$ and $k$ remaining
degrees, respectively. $q_{k}$ is the normalized distribution of
the remaining degree $q_{k}=(k+1)P(k+1)/\sum_{j} j P(j)$, and
$\sigma_{q}^2=\sum_{k} k^2 q_{k} -[\sum_{k} k q_{k}]^2$.
Interestingly, complex networks can be classified into three
types, having $r < 0$, $r\approx 0$ and $r >0$, called the
dissortative, the neutral, and the assortative network,
respectively \cite{Newman02a}.
Most social networks are assortative as shown in TABLE I,
while the Internet and the protein interaction network are dissortative.
While many {\it in silico} models have been introduced, most of them
are neutral.
Thus it would be interesting to introduce an {\it in silico} model
having the assortativity coefficient as positive and large as
empirical values, which would enable one to understand a basic
mechanism of social network formation. We will show that such
assortative networks can be generated via the $q$-component static
model.

\begin{table}[t]
\begin{tabular}{c|ccccc}
\hline\hline
 Name & $N$ & $\langle k \rangle$ & $d$ & $r$ & Ref.\ \\
\hline
 cond-mat & ~~~16,264~ &~~5.85 & ~~6.628~~ & ~~0.185~~ & \cite{Newman01a}\\
 arXiv.org & ~~~52,909~ &~~9.27  &  6.188 & 0.363 & \cite{Newman01a}\\
 Mathematics & ~~~78,835~ &~~4.16 &  8.455 & 0.672 & \cite{Barabasi02a}\\
 Neuroscience & ~~205,202~ &~11.79 &  5.532 & 0.604 & \cite{Barabasi02a}\\
 Video movies & ~~~29,824~ &~33.69 &  4.789 & 0.222 & \cite{imdb}\\
 TV miniseries & ~~~33,980~ &~73.04 &  3.845 & 0.379 & \cite{imdb}\\
 TV cable movies~ & ~~117,655~ &~55.48 &  3.796 & 0.135 & \cite{imdb}\\
 TV series  & ~~~79,663~ & 118.44 & 4.595 & 0.529 & \cite{imdb}\\
\hline\hline
\end{tabular}
\caption{The size $N$, the mean degree $\langle k \rangle$, the diameter $d$,
and the assortativity coefficient $r$ for a number of social
networks.}
\end{table}

\begin{table}[b]
\begin{tabular}{|c|c|c|c|c|c|}
\hline
$N$ & $m$ & $q$ & $d$ & $r$ & ~~similar network~~\\
\hline
~~16,000~~ & ~~2~~ & ~~3,200~~ & ~~6.598~~ & ~~0.174~~ & ~~cond-mat~~\\
~~30,000~~ & ~~5~~ & ~~2,000~~ & ~~4.550~~ & ~~0.218~~ & ~~Video movies~~\\
\hline
\end{tabular}
\caption{Typical simulation results of the diameter $d$ and the
assortative coefficient $r$ obtained under selected conditions of $N$, $m$ and
$q$ with $f=0.2$.}
\end{table}

\section{Model}
The $q$-component static model network is constructed as follows.
Initially, $N$ vertices are present in the system, representing
$N$ people in a society. Each vertex is assigned a $q$-component
weight $(w_{i}^{(1)}, w_{i}^{(2)},\dots,w_{i}^{(q)})$, where $i$
is the vertex index. $w_{i}^{(\mu)}$, the $\mu$-th weight of a
vertex $i$, is given as $\ell_{i,\mu}^{-\alpha_{\mu}}$, where
$\ell_{i,\mu}$ is the rank of the vertex $i$ in the $\mu$-th
subgroup. We take $\{\ell_{1,\mu},\dots,\ell_{N,\mu}\}$ be a
random permutation of the integers $\{1,\dots,N\}$. $\alpha_{\mu}$
is also taken to be a real random number distributed uniformly in
the range $[0.5,1)$. In general, ranks of a person for different
subgroups should be correlated in real society; however, we take
them as independent in this work for simplicity. As the number of
people $N$ becomes large, the number of distinct subgroups $q$ can
increase in real world. Thus, we set $q$ to be $q=sN$ with $s \ll
1$. Then $1/s$ is the average number of people belonging to one
subgroup.

Edges are connected as follows: First, among the $q$ components,
we choose a component $\mu$ at random. Second, we choose two
vertices $(i,j)$ with probabilities equal to normalized weights,
$p_{i}^{(\mu)}\equiv w_{i}^{(\mu)}/\sum_{k} w_{k}^{(\mu)}$ and
$p_{j}^{(\mu)}\equiv w_{j}^{(\mu)}/\sum_{k} w_{k}^{(\mu)}$, and
attach an edge between them with the $\mu$-th color unless an edge
in that color exists already. Edge color is distinct for each
component. Note that the pair $(i,j)$ can be connected via more
than one edge in different colors. Edges in different component
are distinguished by their own colors. The process of attaching
such edges is repeated until $(1-f)mN$ edges are added to the
system. $f$ is a parameter between 0 and 1. We will see that $m$
is related to the average degree. Since the component was chosen
randomly, the number of edges in one color is $(1-f)mN/q$ on
average.

\begin{figure}
\includegraphics[angle=270, scale=0.3]{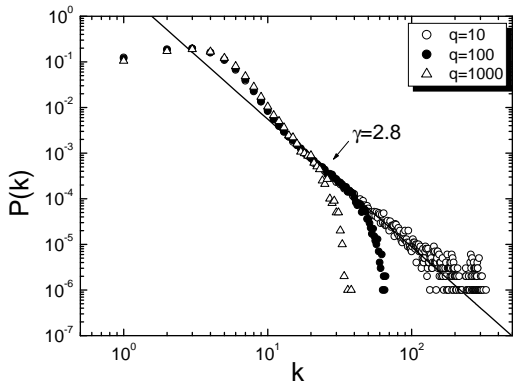}
\caption{The degree distribution $P(k)$ vs the degree $k$ obtained
with $N=10000$ and $m=2$ for $q=10$, 100 and 1000.}
\label{FIG2}
\end{figure}

\begin{figure}
\includegraphics[angle=270, scale=0.3]{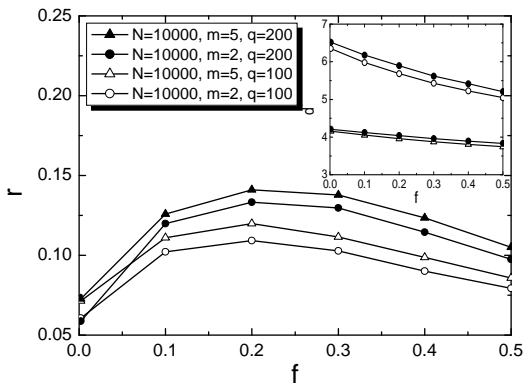}
\caption{The assortativity coefficient $r$ vs the parameter $f$
for various $m$ and $q$. Inset: $f$ dependence of the diameter.}
\label{FIG3}
\end{figure}

To construct a minimal model mimicking social relations, we need
elements playing the role of ``weak ties" \cite{Granovetter73}.
So, to take into account of social relationships among people
having different backgrounds, we suppose that additional social
relationships are formed following the maximum weights among the
$q$ components each individual has. Let
$w_{i,m}=\textrm{max}(p_{i}^{(1)}, \dots, p_{i}^{(q)})$ be the
normalized maximum weight of vertex $i$. Then two distinct
vertices $i$ and $j$ are chosen with probabilities,
$w_{i,m}/\sum_{k} w_{k,m}$ and $w_{j,m}/\sum_{k} w_{k,m}$,
respectively, and are linked by a new color different from the
previous $q$ colors unless such an edge exists already. This
process is repeated until $fmN$ edges are formed. Edges formed by
such maximum weights can be regarded as weak ties, introduced by
Granovetter \cite{Granovetter73} which play an important role in
social networks, connecting different subgroups. We find that the
assortativity coefficient is enhanced by the presence of such weak
ties.

Networks constructed in this way have $mN$ edges with $(q+1)$
colors representing internal structure of subgroups. So, some pair
of vertices are linked by more than one edges in different colors,
albeit such incidences are not so frequent when $q$ is large.
However, when we measure the network properties such as the
shortest pathways, the degree of a vertex, the assortativity
coefficient, and so on, we regard those multiple edges as a single
one. Thus the mean degree $\langle k \rangle$ is slightly less
than $2m$ by about 5\% for typical networks we consider below. A
small size network constructed in this way is shown in FIG.~1.

\section{Simulation results}
We perform numerical simulations
for various values of $q$, $f$, $m$ and $N$, and examine the
diameter $d$ and the assortativity coefficient $r$ as
functions of those parameters. Here the diameter is the average
distance between a connected pair of vertices along the shortest
pathways.

First, the shape of the degree distribution depends on the number
of subgroups. For small $q$, the degree distribution follows a
power law with $\gamma \approx 2.8$, however, for large $q$, it
has a highly skewed form, approximately obeying a power law for a
part of its range, and having an apparently exponential cutoff for
larger $k$. The exponential cutoff for large $k$ originates from
the randomness of ranks of each subgroup, and the SF behavior for
intermediate $k$ does from the SF behavior of each subgroup. A
similar crossover behavior can be found in real social networks,
for example, the collaboration networks of physicists, biologists
and movie stars \cite{Newman02b}.

Second, we examine the assortative coefficient $r$ as a function
of $f$ for a fixed $N$ and several values of $m$ and $q$. As shown
in FIG.~3, the assortativity coefficient exhibits a peak around $f
\approx 0.2$, meaning that the connections among subgroups are
mostly optimized. Thus we limit our further consideration to the
case $f=0.2$. Meanwhile, the diameter gradually decreases with
increasing $f$ as shown in the inset of FIG.~3.

Third, we study the assortativity coefficient $r$ as a function of
$N$ for various $m$ and $q$. It is likely that $r$ increases with
increasing $N$ as $r\sim \ln \ln N$ as shown in Fig.~4, but it
seems to saturate for larger $N$. It also increases with increasing
$m$ and $q$, as shown in FIG.~4. Thus the $q$-component static
model exhibits $r$ values as large as empirical values listed in
TABLE I. Some numerical results of $r$ listed in TABLE II show a
quantitative agreement with the ones obtained in real social
networks.

Fourth, the diameter $d$ is investigated as a function of the
number of vertices $N$ for various $m$ and $q$, as shown in
FIG.~5. It is found that the diameter is proportional to $d \sim
\ln N$ as in the case of random graph, in which $d\sim \ln N/\ln
\langle k \rangle$~\cite{Chung01}.
Furthermore, the diameter becomes smaller as $m$ increases, which
is like the case of random graphs. However, the diameter is almost
insensitive to $q$. To test the so-called ``six degrees of
separation", we extrapolate the straight line in the
semi-logarithmic plot of FIG.~5 to large $N$ for $m=10$ and
$q=N/100$. We obtain $d\approx 6.0$ for $N=10^8$ and $d \approx
6.7$ for $N=10^9$, in reasonable agreement with the Milgram's
``six degrees of separation"~\cite{Milgram67}. Here the choice of
$m=10$ and $q=N/100$ comes from the facts that a person knows
about 20 other people on average (see Chapter 5 of
Ref.~\cite{Barabasi02b}), and there are about 100 members on
average in a subgroup~\cite{Simmel02}.

\begin{figure}
\includegraphics[angle=270, scale=0.3]{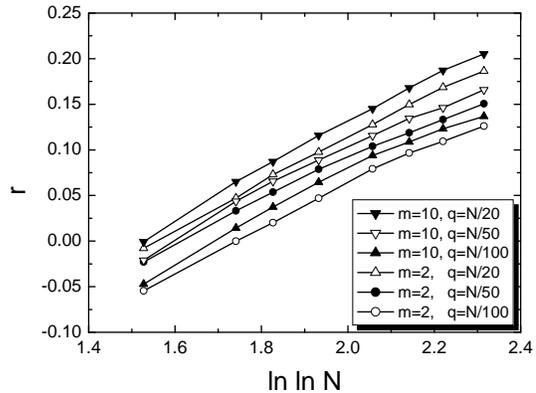}
\caption{The assortativity coefficient $r$ vs $\ln \ln N$ for
various $m$ and $q$ values.}
\label{FIG4}
\end{figure}

\begin{figure}
\includegraphics[angle=270, scale=0.3]{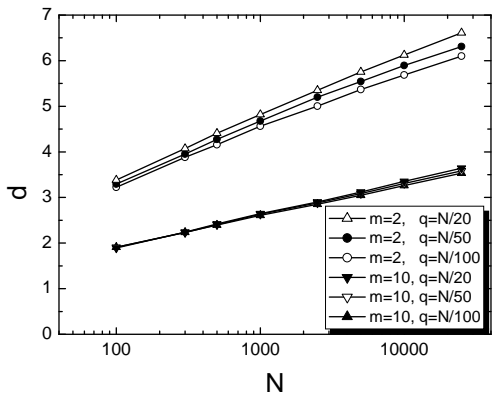}
\caption{The diameter $d$ vs the size $N$ for the same parameters
used in FIG. 4.} \label{FIG5}
\end{figure}

Fifth, one of the properties well studied for complex networks is
the clustering coefficient $C$, which is defined as the average
over all vertices of the ratio of the number of triangles
connected to a given vertex to the number of triples centered on
that vertex. It is known that for the neutral network, the
clustering coefficient behaves as $C(N)\sim
N^{(7-3\gamma)/(\gamma-1)}$ \cite{Newman03a,Newman03c}. Thus when
$\gamma=3$, $C(N)\sim N^{-1}$. For the $q$-component static model,
while $r$ is not close to zero, the rule of attaching edges is such
that no explicit degree-degree correlation enters, so
that it is natural to expect the behavior $C(N)\sim N^{-1}$.
Indeed, the measured behavior is close to the expected one as
shown in FIG.~6, but the slope in FIG.~6 exhibits small deviations
for smaller $q$. For neutral networks, it is known that the
clustering coefficient of a vertex with $k$ is almost
independent of degree $k$. Even in our case, we find that indeed
$C(k)$ is independent of $k$ for different $N$ as shown in the
inset of FIG.~6.

\begin{figure}
\includegraphics[angle=270, scale=0.3]{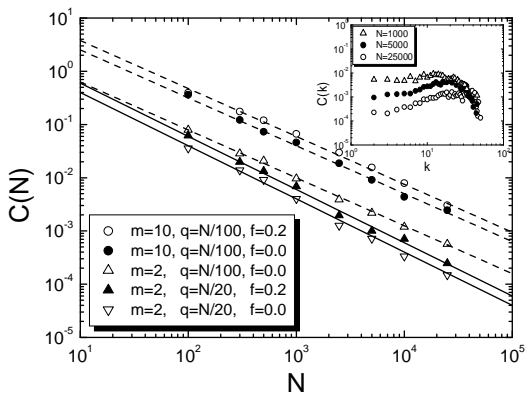}
\caption{Plot of $C(N)$ vs the size $N$ for various $m$, $q$ and
$f$ values. The slopes of the fit lines are $-1.0$ for solid and
-0.9 for the dashed lines, drawn for the eye. Inset : Plot of
$C(k)$ vs degree $k$ for various $N$ with fixed $m=2$, $q=N/20$
and $f=0.2$.} \label{FIG6}
\end{figure}

\section{Conclusions}
We have introduced the $q$-component static model assigning a
$q$-component weight to each vertex. The weight of a given
component of a vertex mimics a weight or fitness of that person in
that subgroup. Through this model, we obtained the diameter of the
acquaintance network as small as the Milgram's ``six degrees of
separation" and the assortativity coefficient as positive and
large as empirical values for a variety of social networks.
Moreover, we obtain the degree distribution in a skewed form,
which is also similar to those of real world social networks. The
clustering coefficients $C(N)$ and $C(k)$ behave as those of a
neutral network, being due to the absence of intrinsic
degree-degree correlation. Such deficiency of the present model
may be improved by introducing the hierarchical structure among
subgroups, or correlated ranks of a person for different
subgroups.

\begin{acknowledgments}
The authors would like to thank K.-I. Goh for sharing his code for
the static model and valuable discussions. This work is supported
by the KOSEF Grant No. R14-2002-059-01000-0 in the ABRL program
and BK21 program of Ministry of Education, Korea.
\end{acknowledgments}

\end{document}